# Layer-dependent Electrocatalysis of MoS$_2$ for Hydrogen Evolution

Yifei Yu[1][§], Shengyang Huang[1][§], Yanpeng Li[1], Stephan Steinmann[3], Weitao Yang[3], Linyou Cao[1,2]*

[1]Department of Materials Science and Engineering, North Carolina State University, Raleigh NC 27695; [2]Department of Physics, North Carolina State University, Raleigh NC 27695; [3]Department of Chemistry, Duke University, Durham, NC 27708

[§] These authors contribute equally.

*To whom correspondence should be addressed.

E-mail: lcao2@ncsu.edu

**Abstract**

The quantitative correlation of the catalytic activity with microscopic structure of heterogeneous catalysts is a major challenge for the field of catalysis science. It requests synergistic capabilities to tailor the structure with atomic scale precision and to control the catalytic reaction to proceed through well-defined pathways. Here we leverage on the controlled growth of MoS$_2$ atomically thin films to demonstrate that the catalytic activity of MoS$_2$ for the hydrogen evolution reaction decreases by a factor of ~4.47 for the addition of every one more layer. Similar layer dependence is also found in edge-riched MoS$_2$ pyramid platelets. This layer-dependent electrocatalysis can be correlated to the hopping of electrons in the vertical direction of MoS$_2$ layers over an interlayer potential barrier, which is found to be 0.119V and consistent with theoretical calculations. Our results point out that increasing the hopping efficiency of electrons in the vertical direction is a key for the development of high-efficiency two-dimensional material catalysts.

Seeking high-efficiency, cost-effective catalysts for mass production of hydrogen gas is critical for the utilization of hydrogen energy. The existing catalysts of Pt-group metals for the hydrogen evolution reaction (HER), which are highly efficient, are too expensive and rare to be useful for the mass production. Molybdenum disulfide ($MoS_2$) promises an earth-abundant, low-cost alternative to the precious metals [1]. Considerable efforts have been dedicated to investigating and optimizing the catalytic activities of various $MoS_2$ materials, including nanoparticles [2-4], nanopores [5], nanowires [6], amorphous and doped $MoS_2$ [7-9], thin films [10], $MoS_2$/graphene heterostructures [11], and chemically exfoliated $MoS_2$ layers [12-15]. It was suggested that the edge site and metallic 1T polymorph of $MoS_2$ materials are catalytically active [2, 14-17]. However, despite the remarkable progress, the rational design of $MoS_2$ structures with optimal catalytic activities has remained elusive due to limited quantitative understanding on the correlation between the catalytic activity and microscopic structure of $MoS_2$ materials.

To quantitatively correlate the catalytic activity with microscopic structure requests capabilities to tailor the structure with atomic scale precision and to control the catalytic reaction to proceed in well-defined pathways. Most of the $MoS_2$ materials studied to date involve a wide dispersion in size, morphology, surface, and crystalline structure. This makes it difficult to assess the role of structural parameters in the catalytic performance. Here we demonstrate a layer-dependent electrocatalysis of $MoS_2$ for the HER by leveraging on a unique synthetic capability that can grow large-area, uniform, and high quality $MoS_2$ atomically thin films with precisely controlled layer numbers. The well-defined physical feature make the film an ideal platform for studies of the structure-catalysis correlation. We find that the catalytic activity, indicated by exchange current density, of the $MoS_2$ film decreases by a factor of about 4.47 for the addition of every one more layer. Similar layer dependence is also found in the electrocatalysis of edge-riched $MoS_2$ pyramid platelets. This layer-dependent electrocatalysis can be correlated to the hopping of electrons in the vertical direction of $MoS_2$ layers over an interlayer potential barrier, which is found to be 0.119V and consistent with theortical calculations.

We grew $MoS_2$ thin films on glassy carbon substrates following a chemical vapor deposition (CVD) process that we previously developed [18]. The layer number was controlled by control of the amount of precursors ($MoCl_5$)[18]. The composition of the film was confirmed as $MoS_2$ by x-ray photoelectron spectroscopy (XPS) characterizations (Fig. S1). The synthesized films are continuous, uniform, and of high crystalline quality, similar to the films we previously grew on sapphire substrates [18]. It can be seen uniformly covering the entire substrate under optical microscopes (Fig. 1a and Fig. S2). Atomic force microscope (AFM) characterizations identify the synthesized monolayer, bilayer, and trilayer $MoS_2$ films in thickness of 0.90 nm, 1.70 nm and 2.15 nm, respectively (Fig. 1b and Fig. S3). These measured results are slightly larger than those of the $MoS_2$ films grown on sapphire [18], which may be related with the relatively larger surface roughness of the glassy carbon substrate (~ 1 nm) than that of sapphire. Raman measurements can further confirm the layer number. The frequency differences ($\Delta k$) of the two characteristic Raman modes ($A_{1g}$ and $E^1_{2g}$ modes), which is widely used to identify the layer number [19], of the synthesized monolayer, bilayer, and trilayer $MoS_2$ films are 20.5, 22.4, and 23 cm$^{-1}$ (Fig. 1c), respectively, consistent with previous results [18, 19]. The AFM and Raman characterizations also confirm the large-area continuity and uniformity of the synthesized films. No substantial voids, steps, and edges were found in the films through numerous AFM measurements. And the frequency difference $\Delta k$ remains reasonably constant during extensive Raman characterizations across the films (Fig. S4). Additionally, the crystalline quality of the synthesized films is reasonably high, as evidenced by the full width at half maximum (FWHM) of the $E^1_{2g}$ mode that is known related with crystalline quality[19]. The $E^1_{2g}$ FWHM of these $MoS_2$ films is comparable with the $MoS_2$ films grown on sapphire, which we previously demonstrated in high crystalline quality by electron microscope and electrical/optical characterizations (Fig.S5) [18].

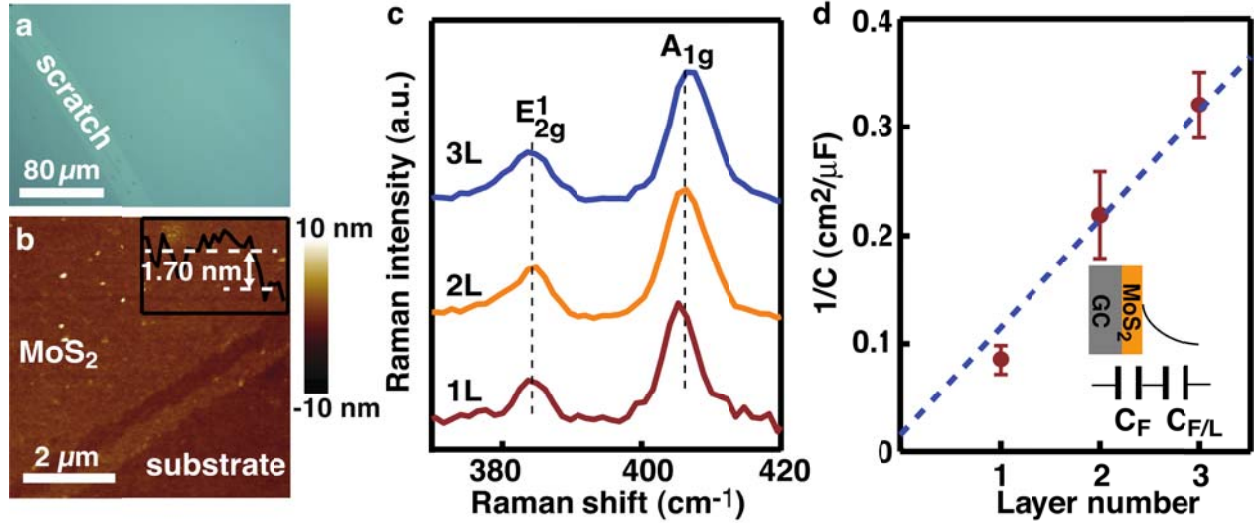

**Fig.1. Structure characterizations of the synthesized $MoS_2$ films.** (a) Optical image and (b) AFM image of a bilayer $MoS_2$ film (monolayer and trilayer see Fig. S1 and S3). A scratch is intentionally made in the film to show the contrast between the substrate and the film and for the convenience of characterizing the height of the film. The inset of (b) is a typical height profile of the film. (c) Raman spectra of the synthesized monolayer (1L), bilayer (2L), and trilayer (3L) $MoS_2$ films. The dashed lines serve to visualize the layer-dependent Raman shift. The characteristic Raman modes are labeled as shown. (d) The reciprocal of measured capacitances of the films as a function of the layer number. Inset is a schematic illustration for the model of two capacitors in series.

The layer number of the film can be confirmed by capacitance measurements . The reciprocal of the measured capacitance ($1/C$) of the film is found linearly dependent on the layer number (Fig. 1d). This linear dependence can be accounted by a model of two capacitors in series, the film with a capacitance of $C_F$ and the double-layer at the film/liquid interface with a capacitance of $C_{F/L}$, $1/C = 1/C_F + 1/C_{F/L}$ (Fig.1d inset). $C_F$ is dependent on the thickness $d$ and dielectric constant of the $MoS_2$ film $\varepsilon_r$, $C_F = \varepsilon_r\varepsilon_0/d$ ($\varepsilon_0$ is the vacuum permittivity). By linearly fitting the measured capacitance with the model, we can find out the double-layer capacitance $C_{F/L}$ as 66.7 µF/cm$^2$, similar to the typical double-layer capacitances previously reported[5]. We can also find the capacitance of monolayer $MoS_2$ $C_F$ as 10 µF/cm$^2$. This suggests the dielectric constant of $MoS_2$ monolayers $\varepsilon_r$ to be 7.63, consistent with the result in reference [20].

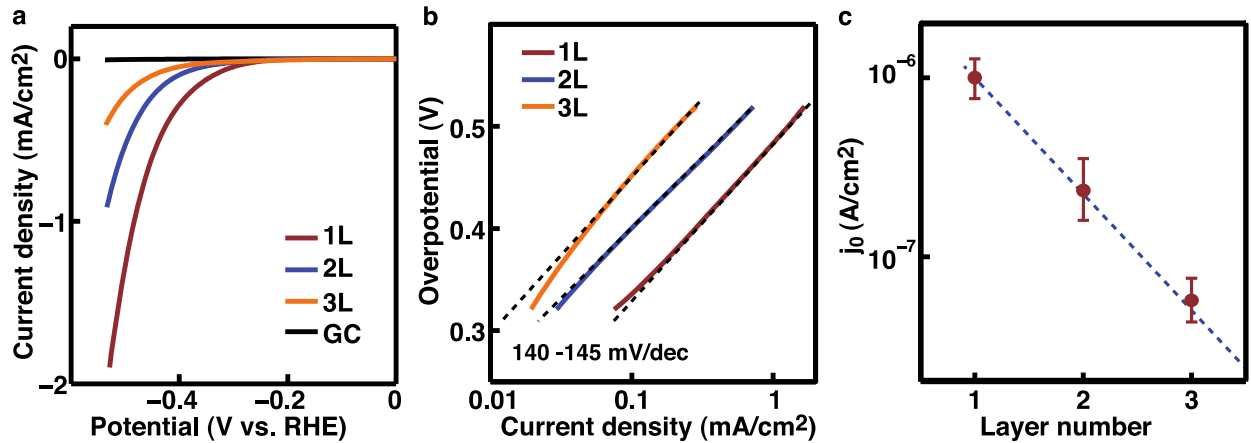

**Fig. 2. Layer dependence of the catalytic activities of MoS$_2$ films.** (a) Polarization curves of the synthesized monolayer (red, 1L), bilayer (blue, 2L), and trilayer (orange, 3L) MoS$_2$ films. The curve of bare glass carbon substrates is also given (black). (b) Tafel plots of the MoS$_2$ films. The dashed lines areee linear fitting for the plots. (c) The exchange current density of the MoS$_2$ film as a function of the layer number. The current density is plotted in a logarithmic scale. The dashed line is a fitting of log$y$ = -0.65$x$-5.35.

The electrocatalysis of the MoS$_2$ film strongly depends on the layer number. Fig. 2a-b shows the polarization curves and corresponding Tafel plots of the synthesized monolayer, bilayer, and trilayer films. The polarization curve of bare glassy carbon electrodes is also given in Fig. 2a, whose trivial current indicates that the glassy carbon is catalytically inactive for the HER. From these results we can find that the cathodic current of the film monotonically decreases with the layer number increasing.

The observed layer dependence can be linked to exchange current densities. By fitting the Tafel plots to the equation of $\eta = \rho\log j + \log j_0$ (overpotential $\eta$, current density $j$, and exchange current density $j_0$), we can find that all the films have a similar slope $\rho$ in the range of 140 – 145 mV/decade, while the exchange current density $j_0$ substantially decreases with the layer number increasing. We repeated the electrocatalytic characterization with numerous MoS$_2$ films (>15 films in total). The results reproducibly show a constant $\rho$ in the range of 140-145 mV/decade and a layer-dependent $j_0$ (Fig. 2c). Interestingly, $j_0$ decreases by a factor of ~ 4.47 for the addition of every one more layer. Note that the observed Tafel slope is larger than typical results on MoS$_2$ materials. This may be related with the high temperature (850°C) used in the synthetic process. Other groups reported a slope of ~ 120mV/decade in the MoS$_2$ materials processed at a temperature of ~500 °C [10, 14, 15], and it has been well known that a processing of MoS$_2$ materials at a higher temperature can result in a higher Tafel slope [10, 14-16, 21]. As will be shown later (Fig. 4), similar Tafel slopes (~140 mV/decade) can be found in the other MoS$_2$ structures (i.e. edge-riched pyramid platelets) synthesized at the same temperature (850°C). This indicates that the Tafel slope is not related with the structure of the film that is of our major interest in this work.

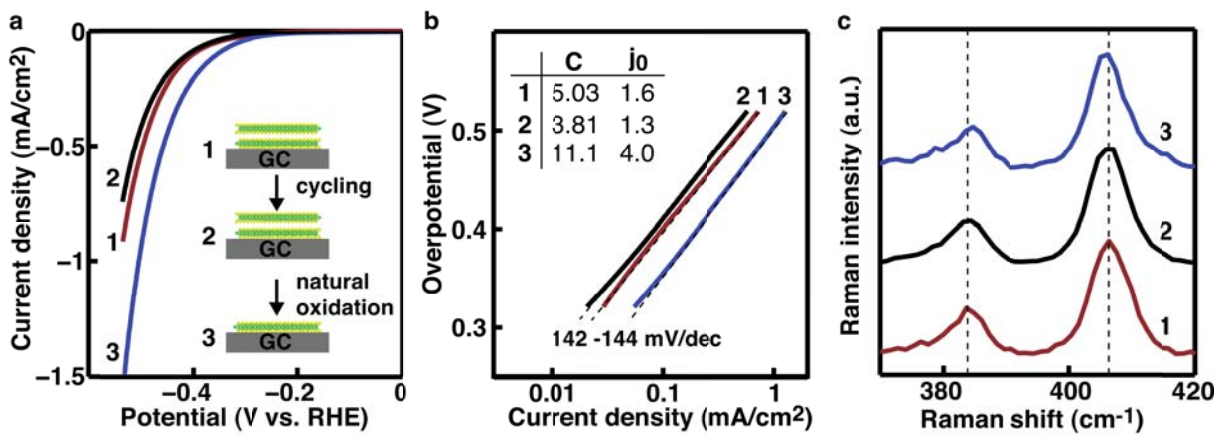

**Fig. 3. Improvement in the electrocatalysis of MoS$_2$ films after the stripping process.** (a) Polarization curves of a bilayer MoS$_2$ film recorded at the different stages in the stripping process: 1, as-grown, 2. after cyclic voltammetry, and 3. after oxidation and dissolution. Inset, schematic illustration for the stripping process. (b) Tafel plots corresponding to the results in (a). The dashed lines indicate the linear fitting for the slope. Inset lists the capacitance ($C$, µF/cm$^2$) and exchange current density ($j_0$, 10$^{-7}$ A/cm$^2$) of the bilayer film at the different stages. (c) Raman spectra of the bilayer MoS$_2$ film at the different stages. The dashed lines serve to visualize the change in the Raman spectra.

The layer-dependent electrocatalysis can be confirmed by a substantial improvement in the catalytic activities of bilayer and trilayer MoS$_2$ films after a stripping process. The stripping process is designed to selectively strip off the outmost layer of the film (Fig. 3 and Fig. S6). It involves procedures of cyclic voltammetry (CV), natural oxidation, and dissolution. We first performed multiple (> 1000 times) CV with the film in a potential range (+0.2 to -0.3V) smaller than the range for the electrocatalysis measurement (+0.2 to -0.6V), and then naturally exposed the CV-processed film to air for weeks (typically one month), which was followed by an immersion of the film into electrolyte solutions. We believe that the outmost layer of the CV-processed film can be gradually oxidized into Mo$^{6+}$ (for instance, MoO$_3$) by the exposure to air, and that the oxidized layer can be dissolved in aqueous solutions to expose the layer underneath due to a weak solubility of MoO$_3$ in water. This stripping mechanism is confirmed by our XPS characterizations through the process. The XPS characterization shows no oxidation (Mo$^{6+}$) in the film right after the CV processing, but substantial oxidation (Mo$^{6+}$) can be found after the CV-processed film being exposed to air for one month. The oxidation (Mo$^{6+}$) peak can be found disappearing again after the oxidized film being immersed into the electrolyte solution (Fig. S7). This indicates that the CV processing does not oxidize the film but may somehow cause it easier to be oxidized by air. The XPS results also indicate that the ratios of Mo$^{6+}$ and Mo$^{4+}$ are around 1:0, 1:1, and 1:2 in the monolayer, bilayer and trilayer films respectively after the natural oxidation (Fig. S8). This suggests that the electrochemical reaction only occurs at the outmost layer, which subsequently limits the oxidation at that layer.

Fig. 3a-b shows the polarization currents and corresponding Tafel plots of a bilayer film recorded through the stripping process (results for the other films see Fig. S6). The current slightly decreases after the CV processing (curve 2 in Fig. 3a) but turns to be much stronger than the initial value after the stripping (curve 3 in Fig. 3a). While the Tafel slope remains pretty

much constant through the process, the exchange current density substantially increases from $1.6 \times 10^{-7}$ to $5 \times 10^{-7}$ A/cm$^2$ after the stripping (Fig. 3b). Capacitance and Raman measurements both confirm that the bilayer film indeed becomes a monolayer after the stripping, the capacitance changing from 5.08 to 11.1 μF/cm$^2$ (Fig. 3b inset) and the frequency Δk from 22.4 cm$^{-1}$ (curve 1 in Fig. 3c) to 20.5 cm$^{-1}$ (curve 3 in Fig. 3c). It is worthwhile to note no obvious change in the Raman spectrum collected right after the CV processing (curve 2 in Fig. 3c). This confirms that the electrochemical process does not cause substantial changes in the structure and composition of the film. Similar increase in the catalytic activity after the stripping process can also be found in trilayer MoS$_2$ films, but understandably not in monolayer MoS$_2$ films (Fig. S6).

It is important to understand the mechanism underlying the observed layer-dependent exchange current density. We can reasonably exclude differences in the composition (i.e. stoichiometry) and crystalline quality (i.e. defects) of the MoS$_2$ films, which might cause difference in the electrocatalysis, because all of them were grown under highly comparable conditions. Additionally, previous studies show that the exchange current density of MoS$_2$ materials linearly depends on the number of edge sites [16]. However, we can exclude the difference in the number of edge sites as the cause for the layer-dependent exchange current densities. The synthesized MoS$_2$ films are continuous, uniform, and show reasonably high crystalline quality (Fig.1). It is safe to conclude that there is no substantial amount of edge sites in the synthesized films and the observed layer dependence is related with the MoS$_2$ atoms in the basal plane.

To further support that the number of edge sites is not the cause, we grew MoS$_2$ pyramid platelets on glassy carbon substrates using temperatures similar to what were used for the synthesis of the film and tested the electrocatalytic performance of the platelets. The synthesized platelets show a monodisperse distribution in size (Fig. 4a lower inset). Different from the film, in which no obvious edges can be found, the pyramid platelet consists of a large number of gradually-shrinking monolayers stacking in the vertical direction with a rich amount of edge sites exposed (Fig. 4a-b). We can find that the pyramid platelets show a very poor electrocatalytic performance (Fig. 4c). The Tafel slope, ~ 140 mV/decade, is similar to what was observed in the film. As discussed in the preceding text, this is related with the synthetic temperature and is not our interest here. The exchange current density, $0.6 \times 10^{-7}$ A/cm$^2$, is barely comparable to those of trilayer thin films. The poor performance of the edge-riched platelets indicates that the number of edge sites is not the cause for the observed layer dependence. Otherwise, we would expect a high exchange current density in the platelets.

Our experimental results also suggest a layer dependence in the electrocatalysis of the edge-riched platelets. We observed the catalytic performance of the pyramid platelets increasing with the size decreasing (Fig. S9). The ratio between the size and height of the pyramid platelet always remains to be constant around 60 (Fig. S9). This constant size/height ratio can enable comparable number of edge sites in unit area among the platelets in different sizes. Therefore, the observed sized-dependent electrocatalysis of the platelet can be ascribed to a dependence on the layer number (height), instead of the density of edge sites. The layer dependence can account for the poorer performance observed in the platelets with a larger height (Fig. 4c). Only the part closed to the substrate electrode can effectively participate the catalytic reaction.

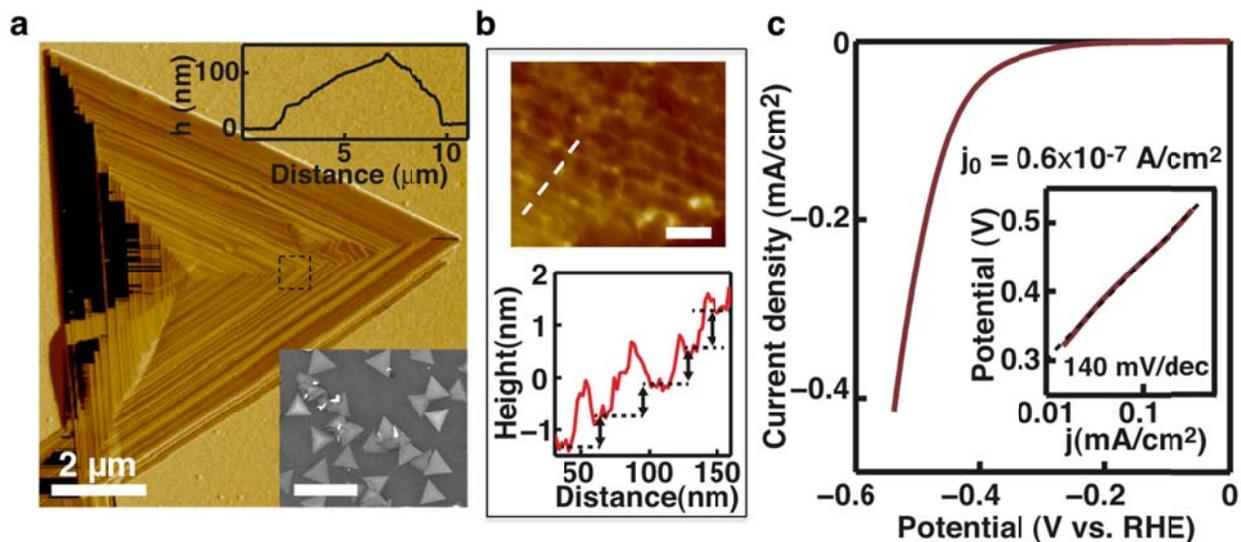

**Fig. 4. Poor electrocatalytic performance of edge-riched MoS$_2$ pyramid platelets.** (a) AFM image of a typical pyramid. Low inset, SEM image of the platelets on glass carbon substrates with a scale bar of 20 μm; upper inset, a typical height profile for the platelet. (b) A magnified image of the area indicated by the dashed square in (a) as well as the height profile for the given dashed white line. The arrows shown in the height profile indicate that the step is one layer. Scale bar, 100 nm. (c) Polarization curve and Tafel plot of the MoS$_2$ pyramid platelet. The given current density considers is normalized to the area of the platelets by considering the partial coverage (60%) of the platelets on the substrate. The Tafel slope and exchange current density are noted as shown.

We find that the layer-dependent exchange current density can be correlated to the hopping of electrons in the vertical direction of MoS$_2$ layers. As discussed in the preceding text, our results indicate that the electrochemical reaction only occurs at the outmost layer of the film. Electrons have to transfer from the glassy carbon electrode to the outmost layer in order to drive the HER (Fig. 4). It is known that the electron transfer in the direction perpendicular to the basal plane of MoS$_2$ materials is through hopping [22], because potential barriers exist in the interlayer gap. We can build up a quantum-tunneling model for the hopping process (Fig. 4). The hopping efficiency is dictated by the interlayer distance $L$ and the potential barrier $V_o$ as $T = e^{-2kL}$, and $k = (2m_eV_o)^{1/2}/\hbar$, where $m_e$ is the effective mass of electrons and $\hbar$ is the Planck's constant. Our results, the exchange current density decreasing by a factor of 4.47 for the addition of every one more layer, suggest $T = e^{-2kL} = 1/4.47$. By substituting the values of $L$ (0.62 nm) and $m_e$ (0.48$m_o$, $m_o = 9.11 \times 10^{-31}$ kg) in MoS$_2$ atomic films [23], we can get $V_o = 0.119$ V. This value is in excellent consistence with a recent theoretical calculation, which predicts that the hopping potential in the vertical direction of MoS$_2$ layers is 0.123 V [24]. The nice consistence suggests a dominant role of the electron hopping in the layer-dependent electrocatalysis of MoS$_2$.

We have demonstrated a layer-dependent electrocatalysis of MoS$_2$ materials and elucidate that the layer dependence is dictated by the interlayer hopping of electrons. The layer dependence is found in both MoS$_2$ atomically thin films, which has little edge sites, and edge-riched strucrures. Different from conventional wisdom, which think the edge site is catalytically active, our results suggests that the atoms in the basal plan can be active sites for catalysis as well. The reportedly better catalytic performance of the edge sites could be because the edge site provides a better way to transfer the electron than the atoms in the basal plane. Our results suggest that increasing the hopping efficiency of electrons is a key for the rational design of $MoS_2$ materials with optimal catalytic activities. The efficiency of hopping is reportedly determined by the interlayer coupling of electron orbitals [24]. Therefore, strategies that can increase the interlayer coupling, such as intercalation of metal ions or atoms, are expected able to enhance the electrocatalytic performance of $MoS_2$ materials.

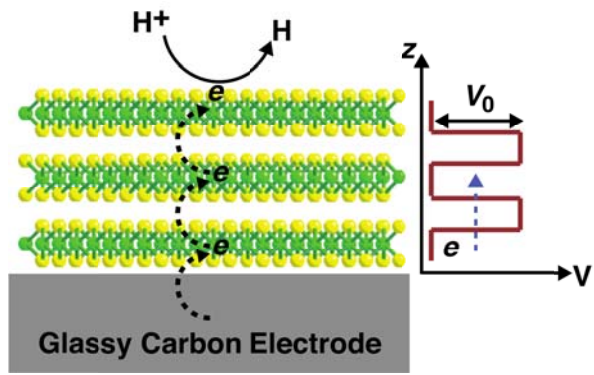

## Methods

*1. Synthesis of $MoS_2$ atomic films with controlled layer number*

$MoS_2$ thin films were synthesized in a tube furnace following a chemical vapor deposition process we have recently developed [18]. In a typical growth, 1-50 mg of molybdenum chloride ($MoCl_5$) powder (99.99%, Sigma-Aldrich) was placed at the center of the furnace and 1g of sulfur powder (Sigma-Aldrich) were placed at the upstream entry of the furnace. Receiving substrates (glassy carbon) were placed in the downstream of the tube. Typical conditions for $MoS_2$ thin film growth included a temperature of 850 °C, a flow rate of 50 sccm, and a pressure around 2 Torr. The precursors were subjected to evaporation at elevated temperatures. The vapor of the precursor materials reacted to produce gaseous $MoS_2$ species, which could subsequently precipitate onto receiving substrates at downstream to yield $MoS_2$ films. The layer number of the synthetic $MoS_2$ thin film can be controlled by control of the amount of MoCl5 precursor. More details can be seen in Ref. 18. $MoS_2$ pyramid platelets were grown with similar conditions as what were used for the growth of thin films. The only difference lies in the pressure. 760 Torr was used for the growth of the pyramid platelets.

*2. Structure and electrochemical characterizations*

The structure and composition of the synthesized films were characterized by tools including optical microscope, atomic force microscope (AFM), Raman spectroscope, and X-ray photoelectron spectroscope (XPS). AFM measurements were performed with a Veeco Dimension-3000 AFM. Raman spectra were collected on a Renishaw-1000 Raman spectroscopy with an excitation wavelength of 514.5 nm. XPS were performed on a SPECS System with PHOIBOS 150 Analyzer using an Mg Kα X-ray source.

The electrochemical characterization of the $MoS_2$ films were performed in 0.5 M $H_2SO_4$ using a CH Instrument electrochemical analyzer (Model CHI604D), a Pt-wire counter electrode, a saturated calomel reference electrode (SCE), and a glassy carbon (0.785 cm$^2$) working electrode. Nitrogen gas was bubbled into the electrolyte throughout the experiment. While all the electrochemical characterization studies were performed using a saturated calomel reference electrode (SCE), the potential values mentioned in the electrochemical study are referred to a reversible hydrogen electrode (RHE). Calibration of the reference electrode for the reversible hydrogen potential was performed using a platinum (Pt) disk as working electrode and a Pt wire as counter electrode in 0.5 M H$_2$SO$_4$. The electrolyte was purged with ultrahigh purity hydrogen (Airgas) during the measurement. The potential shift of the SCE is -0.262 V vs. RHE. The electrocatalysis was measured using linear sweeping from +0.2 V to -0.6V (vs. RHE) with a scan rate of 5 mV/S. The cycling voltamettry was performed in a range of +0.2 to -0.3 V (vs. RHE) at a scan rate of 50 mV/s.

The capacitance of the film was characterized using electrochemical impedance spectroscopy (EIS). The AC impedance is measured within the frequency range of 10$^5$ to 1 Hz with perturbation voltage amplitude of 5 mV. An equivalent Randles circuit model was fit to the data with ZSimpWin software to determine the system resistance and the capacitance.

**Acknowledgement**

L. C. acknowledges a Young Investigator Award from the Army Research Office (W911NF-13-1-0201).

**Author Contributions**

Y. Y. and L. C. conceived the experiments. Y. Y., S. H., and Y. L. perfermed the experiments. Y. Y., S. H., and L.C. analyzed the data. S. S., W. Y., and L. C. came up the mechanism. All authors were involved in writing the manuscript.

**Competing financial interests**

The authors declare no competing financial interests.

# Layer-dependent Electrocatalysis of MoS$_2$ for Hydrogen Evolution

Yifei Yu[1][§], Sheng-Yang Huang[1][§], Yanpeng Li[1], Stephan N. Steinmann[3], Weitao Yang[3], Linyou Cao[1,2]*

[1]Department of Materials Science and Engineering, North Carolina State University, Raleigh NC 27695; [2]Department of Physics, North Carolina State University, Raleigh NC 27695; [3]Department of Chemistry, Duke University, Durham, NC 27708

[§] These authors contribute equally.

*To whom correspondence should be addressed.

E-mail: lcao2@ncsu.edu

This PDF document includes

Fig. S1-S9

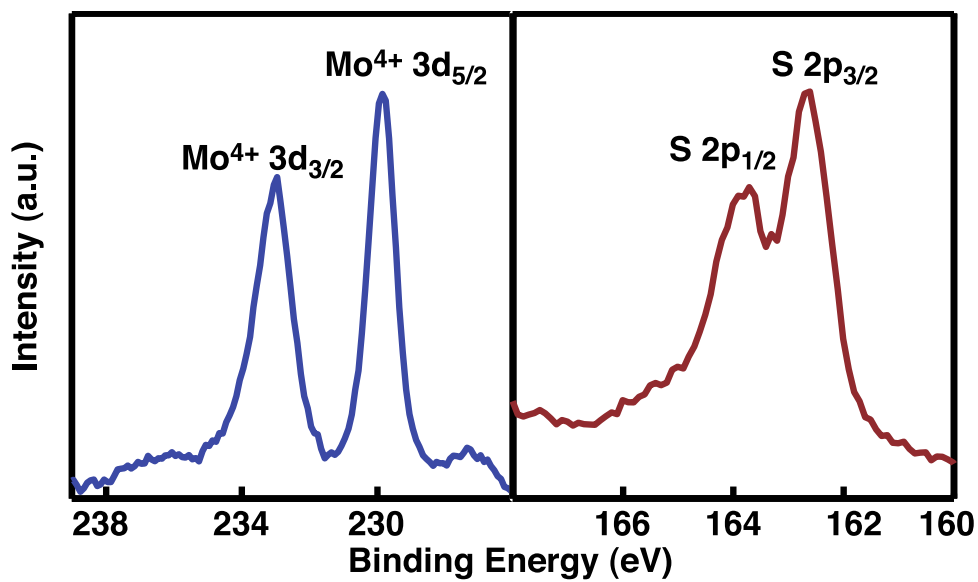

**Fig. S1. XPS of the synthesized MoS$_2$ films.** Binding energies for (**Left**) the Mo atoms and (**Right**) the sulfur atoms. The XPS peaks are assigned as shown.

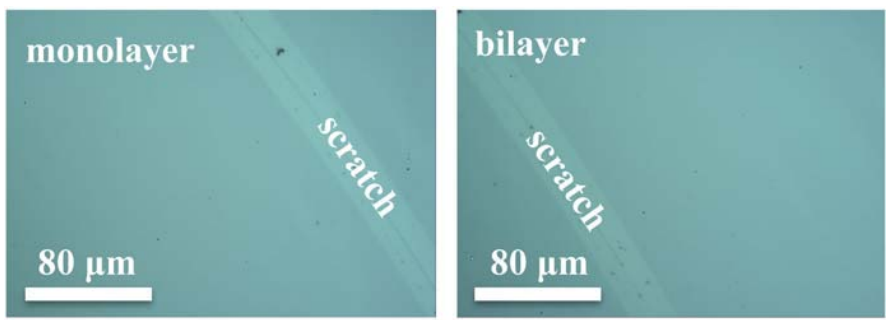

**Fig.S2. Optical images of monolayer and bilayer MoS$_2$ films grown on glassy carbon substrates.** Scratches (brighter regions) are intentionally introduced in the films to show the color contrast between the substrate and the film. We can see that the bilayer is a little bit darker than the monolayer.

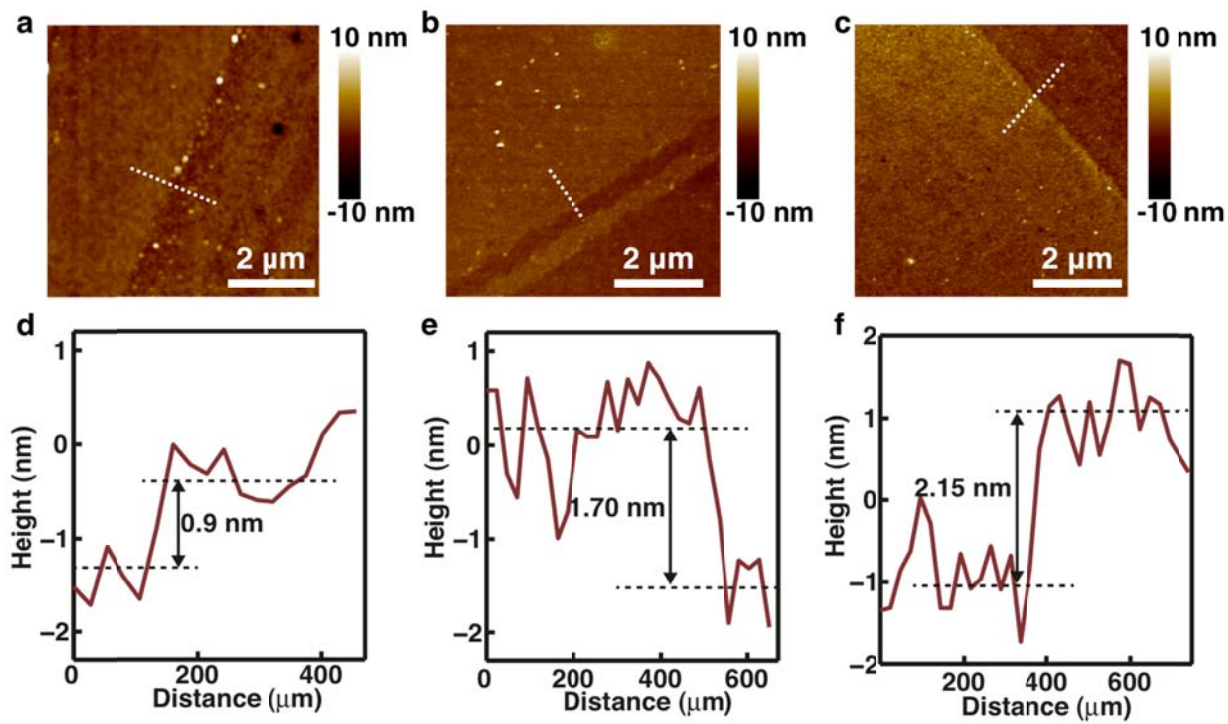

**Fig. S3. AFM characterizations.** (**a-c**) Typical AFM images of the synthesized monolayer, bilayer, and trilayer films grown on glassy carbon substrates. The area occupied by the $MoS_2$ is labeled as shown. (**d-f**) Height profiles for the white dashed lines shown in (a-c), respectively.

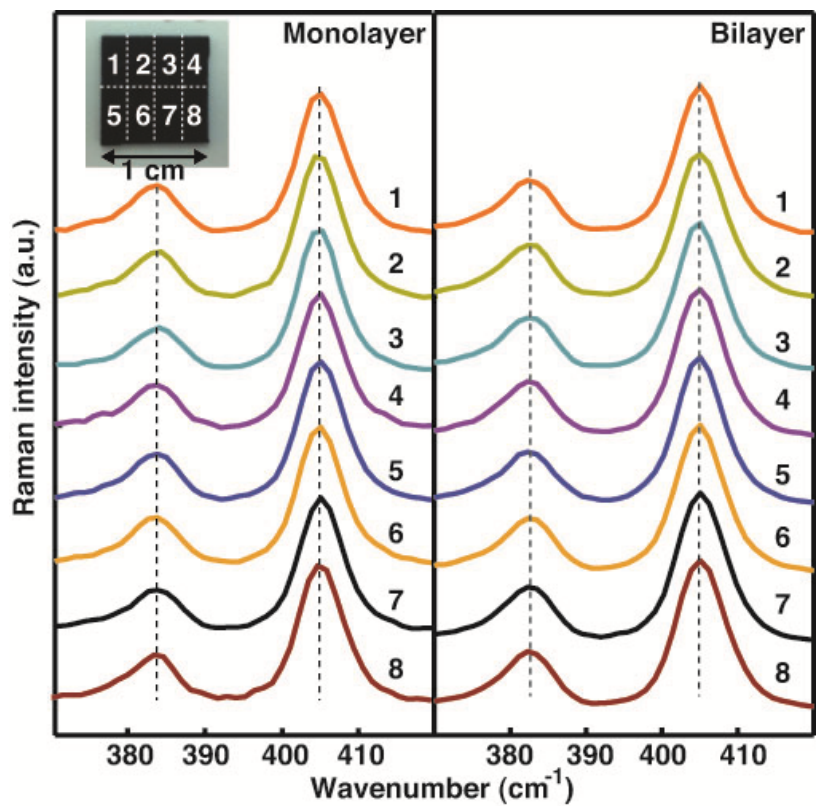

**Fig. S4. Raman spectra collected from numerous locations of the MoS$_2$ film.** Eight of the collected Raman spectra for the monolayer and the bilayer are given as shown. The inset schematically shows the eight spectra collected from locations evenly distributing across the film. The dashed lines serve to visualize the constant Raman peaks across the entire film.

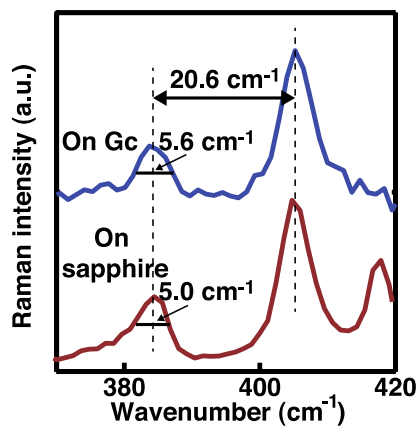

**Fig. S5. Raman spectra of the monolayer MoS$_2$ films grown on the substrates of glassy carbon (GC, blue) and sapphire (red).** The frequency difference between the two characteristic Raman modes of MoS$_2$ is given. Also given is the full width at half magnitude (FWHM) of the $E^1_{2g}$ mode (~ 385 cm$^{-1}$) in the films.

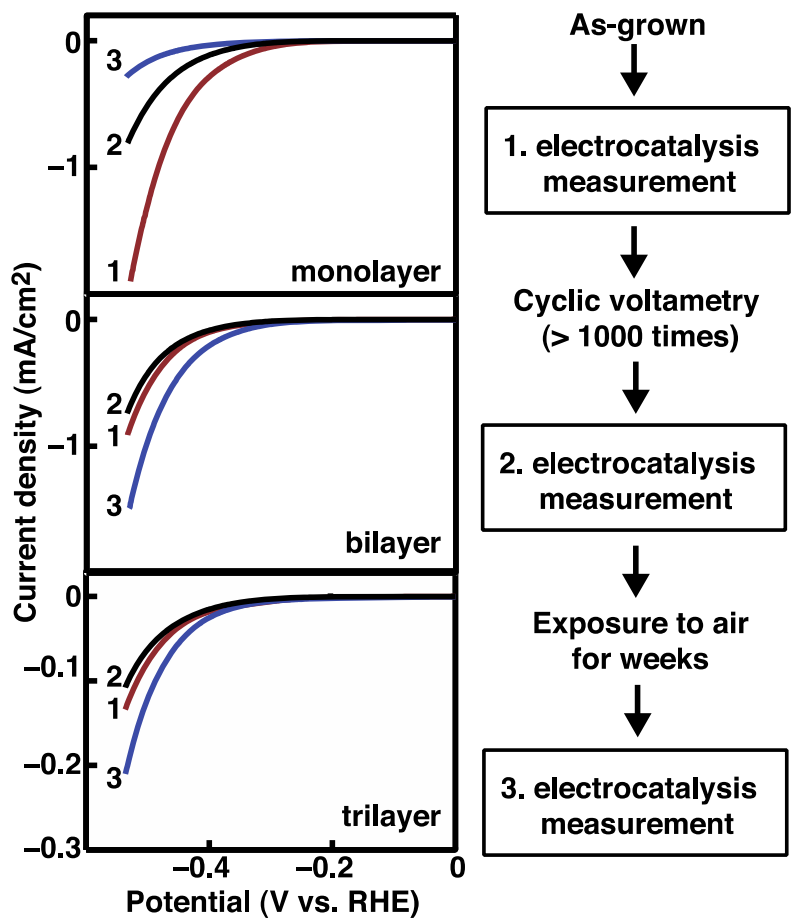

**Fig.S6. Polarization curves of the MoS$_2$ films in the stripping test**. The polarization curves of the monolayer, bilayer, and trilayer MoS2 films were recorded at different stages of the stripping test. The diagram to the right illustrates the different stages, 1. as-grown films, 2, after the cyclic voltammetry processing, and 3. After exposing the CV-process films to air for weeks.

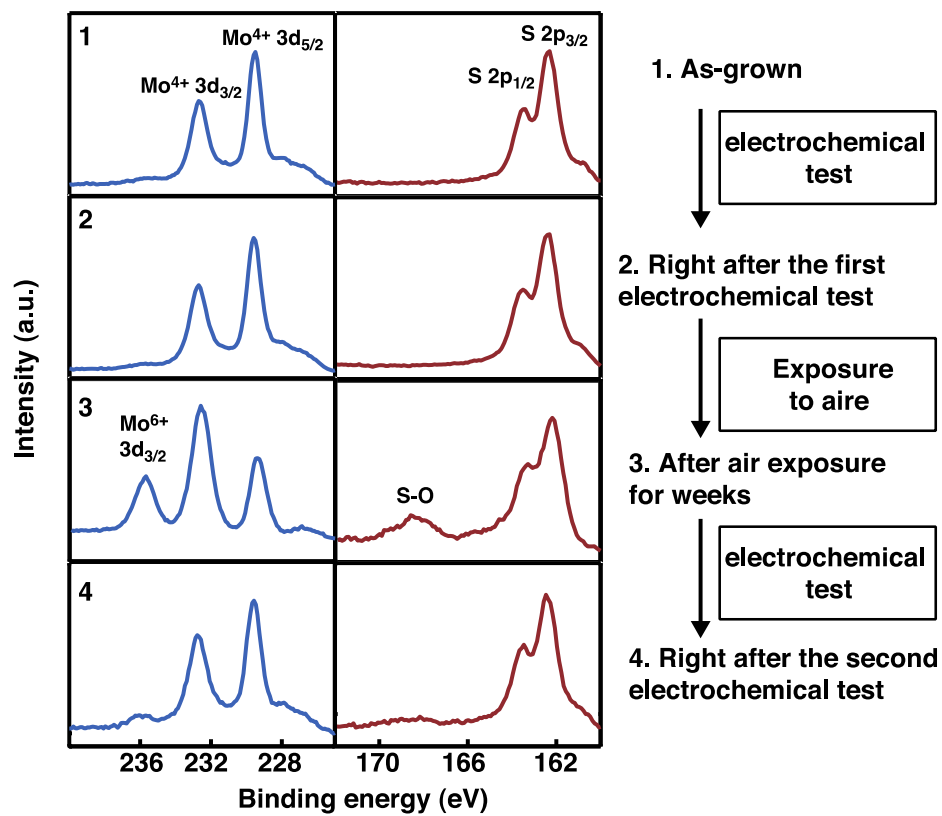

**Fig. S7. XPS characterization of the MoS$_2$ films through the stripping test.** The composition of the MoS2 film was monitored at the different states through the stripping test. Both the binding energy of (**Left**) the Mo atoms and (**Middle**) the sulfur atoms are given. The diagram to the right illustrates the different stages, 1. as-grown films, 2, after the cyclic voltammetry processing, 3. after exposing the CV-process films to air for weeks, and 4, after immersing the air-exposed films into electrolyte solutions.

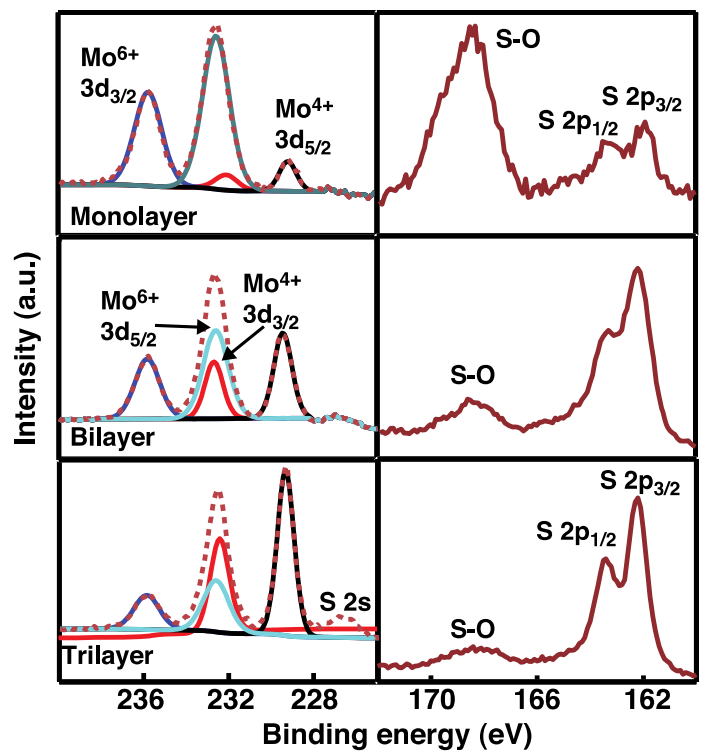

**Fig. S8. XPS characterization of the oxidation in MoS$_2$ films.** The XPS characterizations were performed with the monolayer, bilayer, and trilayer MoS2 film after the CV processing and exposure to air for weeks. Both the binding energy of (**Left**) the Mo atoms and (**Middle**) the sulfur atoms are given. The XPS peaks are assigned as shown. Curve fitting was performed for the results of Mo atoms in order to find out the ratio of Mo$^{6+}$ and Mo$^{4+}$, which is demonstrated as 89.7:10.3, 51.73:48.27, and 33.91:66.09 for the monolayer, bilayer, and trilayer films, respectively.

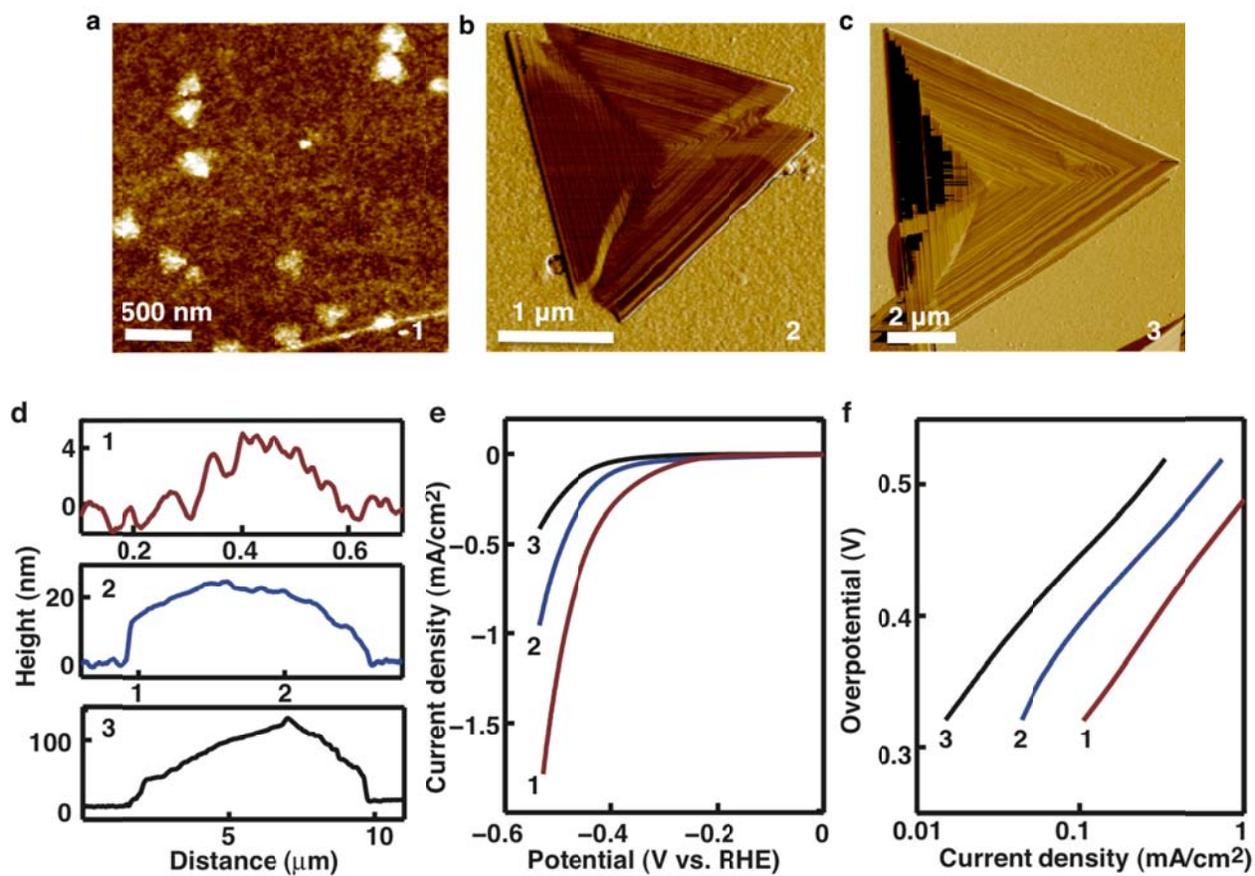

**Fig. S9. Characterizations of $MoS_2$ pyramid platelets grown on glassy carbon substrates.** (**a-c**) AFM images of $MoS_2$ pyramid platelets with different sizes. For the convenience of discussion, each of these platelets is assigned with a number, 1, 2, and 3, respectively. (d) Typical height profile obtained from AFM characterizations of the three types of platelets. (e-f) Typical polarization curves and corresponding Tafel plots of the triangle platelets.